\documentclass{elsart}
\usepackage{amssymb}
\usepackage{epsfig}

\input{tcilatex}
\input tcilatex
\begin{document}
%\pagestyle{myheadings}
%\markboth{FREIBURG-EHEP-97-16}{FREIBURG-EHEP-97-16}

\begin{frontmatter}
\title{AC-coupled GaAs microstrip detectors\\
with a new type of integrated bias resistors}
\author{R. Irsigler$^{1}$, R. Geppert, R. G\"{o}ppert, M. Hornung, J. Ludwig,}
\author{M. Rogalla, K. Runge, Th. Schmid, A. S\"{o}ldner-Rembold,}
\author{M. Webel, C. Weber}
\address{Albert-Ludwigs-Universit\"{a}t Freiburg, Fakult\"{a}t f\"{u}r Physik, \\
D-79104 Freiburg, Germany}

\begin{abstract}
Full size single-sided GaAs microstrip detectors with integrated coupling
capacitors and bias resistors have been fabricated on 3'' substrate wafers.
PECVD deposited SiO$_2$ and SiO$_2$/Si$_3$N$_4$ layers were used to provide
coupling capacitaces of 32.5 pF/cm and 61.6 pF/cm, respectively. The
resistors are made of sputtered CERMET using simple lift of technique. The
sheet resistivity of 78 k$\Omega $/sq. and the thermal coefficient of
resistance of less than 4$\times $10$^{-3}$/${{}^{\circ }}$C satisfy the
demands of small area biasing resistors, working on a wide temperature range.
\end{abstract}
\end{frontmatter}

\footnotetext[1]{
Corresponding author, Tel.: +49 761 203 5911, fax: +49 761 203 5931, e-mail:
irsigler@ruhpb.physik.uni-freiburg.de}

\section{Introduction}

Several aspects on the design of a semiconductor microstrip detectors has to
be taken into account in order to get a good signal to noise ratio. First of
all, the strip capacitance (the sum of the interstrip capacitances between
neighbouring strips and the body capacitance of the strip) should be low
because it determines the noise level of the readout electronics \cite{Bar}.
For short shaping times, there is almost no additional contribution due to
the shot noise of the detector. This is still the case for the higher
leakage currents of GaAs detectors ($\sim $20 nA/mm$^2$) \cite{Mar} compared
to standard Si detectors ($\sim $0.5 nA/mm$^2$)\cite{Pit}. Short shaping
times have to be used because of the very high luminosity (10$^{34}$ cm$%
^{-2} $s$^{-1}$) and high bunch crossing rate (40 MHz) at future high energy
physics experiments like LHC \cite{Pet}.

Secondly, the interstrip capacitance has to be large compared to the body
(backplane) capacitance of the strip to avoid signal losses to ground.

Thirdly, the coupling capacitance has to be magnitudes higher compared to
the strip capacitance in order to avoid a signal spreading to neighbouring
strips \cite{Cac}.

Biasing resistors are necessary to drain out the leakage current of the
detector. In conjunction with the strip capacitors, they act as a low pass
filters. Values in the range of M$\Omega $ are needed to avoid signal losses
to ground.

In addition strip resistance has to be minimized to reduce dispersion of the
signal pulse during penetration at the transmission line of the strip \cite
{Gad}.

Due to limited space resources, it is not possible to integrate coupling
capacitors and bias resistors on VLSI-amplifier chips. Hence external
capacitor and resistor chips have to be used, or they have to be integrated
onto the detector. On integrated detectors, a reduced number of
interconnections have to be made which improves yield and reliability. On
the other hand additional processing steps raise cost and complexity of
detector fabrication.

Simplification of processing steps is an essential task in detector design.
One of the major advantages of GaAs detectors is the fact that simple
Schottky contacts can be used instead of diffused or implanted pn-contacts
in silicon technology. No intermediate p-stops are needed to compensate
accumulated surface charge between n-strips. Although GaAs substrate wafers
are more expensive than Si wafers, a reduced number of masks and processing
steps makes GaAs microstrip detectors competitive to standard Si-detectors.

The design, fabrication and electrical performance of integrated
GaAs-microstrip detectors are described in the following sections.

\section{Wafer design}

The wafer design contains several detectors and test structures. In
accordance to the SCT96 layout specification of the ATLAS\ detector at LHC 
\cite{SCT}, a keystone detector was designed which covers the main part of
the wafer. The 6 cm long detector with 256 strips has a tilt angle of 3$%
^{\circ }$ and a varying pitch from 80 $\mu $m at the top to 68 $\mu $m at
the bottom of the strips. The gap between the strips is constantly 25 $\mu $%
m. At the bottom, each strip is connectet via a biasing resistor to the
common bias bar.

In addition, four detectors with a reduced lenght of 1.5 cm, 50 $\mu $m
pitch and 256 strips were placed on the wafer design. Two of them are
detectors with a variable width of the strips (40/30/25/20 $\mu $m) which
are grouped in 64 strips each. All detectors are AC-coupled. The detectors
are surrounded with some test structures to measure the performance of the
biasing resistors and coupling capacitances.

The mask set consists of six layers. The first one defines the strips, guard
ring, bond pads and biasing line of the detector. The second mask opens the
contact windows for the bond pads and the resistors in the dielectric layer.
The third mask is used for the CERMET\ resistors. The resistor lines lay on
top of the dielectric and are connected to the strips and the common biasing
line via etched holes. The serpentine design was selected because of a
better utilisation of the area. The fourth mask defines the top strip
metallization for the capacitors and provides vias over the guard ring to
the bond pads at the first level metallization. Mask five and six define the
backside metallization of the wafer and opens contact holes in the backside
passivation.

Fig.1 \label{cerres3}shows the integrated CERMET resistors at the end of a
50 $\mu $m pitch microstrip detector. The strips are surrounded by a guard
ring.

\section{Device fabrication}

All detectors have been designed and fabricated in our laboratory at the
Materials Research Center in Freiburg. Detectors and test structures of
various geometries were processed on 3 in. semi insulating GaAs from
Freiberger Compound Materials. Before deposition of the contacts, wafers
were cleaned in acetone and iso-propanol with a subsequent etch in HCl/H$_2$%
O and NH$_4$OH/H$_2$O$_2$/H$_2$0. Either Ti/Pt/Au/Ni (10/20/80/5 nm) or
Ti/Ti-W/Al/Ti-W (10/10/100/5 nm) layers were used as first level
metallization. Depending on the barrier layer, the Schottky contacts are
stable up to process temperatures between 400 ${{}^{\circ }}$C an 500 ${%
{}^{\circ }}$C. RBS measurements have shown that the sputtered Ti-W barrier
layer exhibits a better performance with respect to the temperature
stability of the Schottky contact. In either case thermal budget is a
critical point during detector fabrication. The strip resistance was in the
range between 150 $\Omega /$cm and 200 $\Omega /$cm for 30 $\mu $m wide
Strips.

Single layers of SiO$_2$ or double layers of SiO$_2$/Si$_3$N$_4$ were
deposited at 300${{}^{\circ }}$C in a PECVD process. Afterwards contact
holes were etched into the dielectric layers to provide interconnections to
the resistors and the second level metallization (see Fig. 2). The etch mask
was also used for a Ni/Au/Ni (10/90/10 nm) plug fill of the contact holes.
In the next step, CERMET was sputtered onto the wafer to define the
resistors after the lift off process. Then, the second level metallization
was deposited using either evaporated Ni/Au (10/100 nm) layers or sputtered
Al (120 nm) to provide the coupling capacitors.

Next, the front side was covered with photoresist to protect the surface and
the originally 625 $\mu $m thick wafers were lapped and polished down in a
CMP process to a residual thickness of 200 $\mu $m. The backside of the
wafers were O-implanted at an energy of 130 keV with a dose of 1 $\times $ 10%
$^{13}$ cm$^{-2}$. The resulting damage induced isolation layer improves the
breakdown behavior at full depletion \cite{Irs}. Afterwards, the backside
was patterned photolithographically in a double sided mask aligner to define
the backside contact under the strips. Sputtered Ti-W/Al (10/120 nm) was
used as metallization. Finally a layer of PECVD SiO$_2$ was deposited and
etched to protect the backside from scratches during handling and mounting
of the detectors.

\section{Interstrip capacitance}

Variation of the interstrip capacitance with the gap was measured at the
variable width detector. As expected, the interstrip capacitance is a
decreasing function with the separation of the strips. As shown in Fig.3,
the interstrip capacitance to the first neighbour strip varies between 1.15
pF/cm at 10 $\mu $m separation down to 0.67 pF/cm for a gap of 30 $\mu $m.
The interstrip capacitance to the second neighbour strip is about 60 \% of
this. Hence a total strip capacitance in the range between 2.5 pF/cm and 4.5
pF/cm could be expected for the considered detector geometries. Demanding a
ten times higher coupling capacitance, the coupling capacitance should be in
the range between 150 pF and 270 pF for 6 cm long strips.

\section{Coupling capacitors}

Because of the poor electric properties of native GaAs oxides (As$_2$O$_3$,
Ga$_2$O$_3$), foreign dielectric materials have to be used in GaAs device
processing. Materials available for dielectric layers in microelectronics
are usually SiO$_2$, Si$_3$N$_4$, Al$_2$O$_3$, Ta$_2$O$_5$ and Polyimide 
\cite{How}. All of those materials are showing some strengths and weaknesses.

Depending on the deposition technique and conditions the dielectric constant
is between 4 and 5 for SiO$_2$ and between 6 and 9 for Si$_3$N$_4$. Usually
they are deposited in a PECVD process at relatively low temperatures (300 ${%
{}^{\circ }}$C). Both layers are easily wet etchable in buffered HF to open
contact windows. A combination of both layers are often preferred because of
the lower breakdown voltage due to pinholes formation in SiO$_2$ and the
high intrinsic stress of Si$_3$N$_4$ when single layers are used \cite{Tsa}.

Polyimide can be easily spun onto surfaces like a photoresist but require
high curing temperatures ( \TEXTsymbol{>} 400 ${{}^{\circ }}$C) to achieve
best dielectric properties. Dielectric constant is in the range between 3
and 4. Film thickness is hard to control and patterning has to be done using
dry etching in an oxygen plasma \cite{How}.

Al$_2$O$_3$ and Ta$_2$O$_5$ have high dielectric constants of 9.5 and more
than 20, respectively. Sputtering methods can be used for deposition but
surface damage and variation of film thickness can be significant\cite{Par}.

Within this work, the materials of choice were single layers of SiO$_2$ (300
nm) and double layers of SiO$_2$/Si$_3$N$_4$ (100 nm/200 nm) deposited in a
PECVD process. Some results are presented in the following section.

\subsection{Measurements}

All measurements have been performed on wafer at a probe station using a HP
4284 A LCR-meter in parallel mode. Fixed frequency measurements were done at
10 kHz with a oscillating level of 200 mV and a DC-bias voltage of 2 V in
accordance to the expected voltage drop at the biasing resistors.

A comparison between the different dielectric layers is shown in Fig. 4 for
a detector with variable strip length and in Fig. 5 for a detector with
variable strip width. The calculated dielectric constants for SiO$_2$ and SiO%
$_2$/Si$_3$N$_4$ are 4.0 and 7.3, respectively. For a strip width of 35 mm,
the measured coupling capacitances were 23.0 $\pm $ 0.2 pF/cm and 42.0 $\pm $
0.5 pF/cm, respectively. The full size keystone detector with 80 $\mu $m
pitch, constant gap of 20 $\mu $m, constant length of 6 cm and varying width
from top to bottom of the strip had a coupling capacitance of 200 pF for SiO$%
_2$ and 370 pF for SiO$_2$/Si$_3$N$_4$.

As plotted in Fig. 6, the coupling capacitance is frequency independent over
a wide range. At frequencies above 100 kHz a significant decrease is
observed for the full size keystone detector. This is due to the fact, that
the coupling capacitors have to be treated as a distributed transmission
line of finite resistors and capacitors\cite{Gad}. At high frequencies, the
effective length of the capacitor is reduced, resulting in a lower
capacitance. This is especially the case when implanted strips in
Si-detectors have a high resistivity and were not covered with a
metallization layer \cite{Bar}. For the GaAs detector of shorter lengths and
smaller resistances, this effect is negligible.

\section{Biasing resistors}

Using lift-off technique, film thickness is normally restricted to less than
200 nm. In width, the resistor area is restricted by the pitch of the
strips. The length should be as small as possible because it acts as a dead
part of the detector. On the other hand, structures become more sensitive to
varying deposition conditions if the dimensions are to small. A suitable
compromise is a resistor area in the rage of 50 $\mu $m $\times $ 250 $\mu $%
m. Hence the sheet resistivity of the resistor material must be in the range
of 50 - 100 k$\Omega $/sq.  to achieve resistor values in the M$\Omega $
range.

Microstrip detectors in the ATLAS-experiment at LHC has to operate at a
temperature of -10 ${{}^{\circ }}$C for a period of at least 10 years.
Therefore it is recommended that the resistor material has a low temperature
coefficient..

The requirements of a high sheet resistance, long term stability and weak
temperature dependence limits the suitable alternatives for resistor
materials. Different approaches have been evaluated to integrate biasing
structures on microstrip detectors so far. This includes passive components
like polysilicon resistors \cite{Ohs} as well as active devices like
punch-through biasing \cite{Ell} and FOXFET structures\cite{All}.

Polysilicon has to be deposited at rather high temperatures in a LPCVD
process and local implantation steps with post annealing at high temperature
has to be applied to get a good ohmic junction at the metal
strip/polysilicon interface \cite{Yeh}. Moreover, if pure Al is used as
strip metallization, spiking problems could degrade reliability of the
resistors.

Punch-through biasing needs no extra processing steps but suffers from a
leakage current dependent dynamic resistance which causes considerable base
line differences between channels.

On the FOXFET-structure, a gate electrode covering the punch-through gap
controls the dynamic resistance. Unfortunately no integration of
FOXFET-structures on GaAs detectors is possible because of the well known
pinning of the Fermi level to the middle of the band gap at the
dielectric/GaAs interface \cite{Wie}

This work is focussed on a new type of biasing resistors for microstrip
detectors which will be discussed in the following section.

\subsection{Integrated CERMET-resistors}

Thin film resistors made of CERMET have been widely used in microelectronic
industry for a long time \cite{Ame1} - \cite{Hof}. CERMET (CERamic/METal) is
a mixture of an insulator (SiO) and a metal (Cr or Au). In this two phase
material current transport is interpreted in terms of electron tunneling
between metal islands in the insulator matrix \cite{Mor1}. Simultaneously
evaporation \cite{Mor2} or sputtering \cite{Ste} can be used for deposition.

It was found, that the electrical properties are very sensitive to the
CERMET composition and the deposition conditions (substrate temperature,
sputtering power, post annealing)\cite{Ste,Hof}. Reducing the Cr content in
the sputtering target from 50 $\%$ to 10 $\%$ by volume results in a drastic
increase in sheet resistance from 10$^3$ $\Omega $/sq. to 10$^{13}$ $%
\Omega $/sq. Hence a wide range of resistor values can be obtained
by choosing appropriate target composition.

\subsection{Measurement}

A target composition of 55 Vol. \% SiO / 45 Vol. \% Cr was chosen in order
to reach the mentioned demands on the sheet resistivity. Rf-magnetron
sputtering in DC mode have been performed in a sputtering system from von
Ardenne (LA250).

Fig. 7 shows the I-V characteristics of the integrated CERMET biasing
resistors for two different sputtering powers. In order to achieve a
comparable resistor thickness, the sputtering time has to be increased form
100 seconds at 200 Watt to 250 seconds at 100 Watt because of lower
deposition rates at reduced power. Adhesion of the CERMET\ film was found to
be excellent in any case. The final thicknesses were measured with a stylus
profiler (Tencor P10) giving values of 133 $\mu $m and 121 $\mu $m for 200 W
and 100 W, respectively. From the slope of the I-V curve, the resistance was
calculated to be 4.85 M$\Omega $ at 200 W and 2.47 M$\Omega $ at 100 W
sputtering power.

Some test vehicles were used to measure the resistance as a function of
resistor length. The width of the resistor line was 10 $\mu $m. The
corresponding resistivity was calculated to be 0.95 $\Omega $cm and 3.14 $%
\Omega $cm for 100 W and 200 W sputtering power, respectively. This
corresponds to a sheet resistance of 78.4 k$\Omega $/sq. and 236 k$%
\Omega $/sq., respectively.

In steps of 5 K, I-V curves as a function of temperature between -40 ${%
{}^{\circ }}$C and +60 ${{}^{\circ }}$C were measured in a temperature
controlled chamber. The temperature behavior is described by the thermal
coefficient of resistance (TCR), which is defined as:

\[
\alpha =\frac{\Delta R}{R*\Delta T} 
\]

The TCR $\alpha $ can be calculated from the relative change of resistance $%
\Delta R/R$ due to a temperature change $\Delta T$. A comparison of the TCR
between the integrated CERMET resistors and a external resistor chip%
\footnote{%
3,2 M$\Omega ,$Kharkov, Ucraine} is shown in Fig.8. External R-chips are
frequently used, if no biasing resistors are integrated on the detector. In
both cases the TCR\ is negative and smoothly increasing with temperature.
For the CERMET\ resistors a value of -4 $\times $ $10^{-3}/{{}^{\circ }}$C
at -10 ${{}^{\circ }}$C reaching -2.5 $\times $ $10^{-3}/{{}^{\circ }}$C at
room temperature was calculated. Those TCR values are even lower than the
corresponding values for the external resistors and show the good
performance of the CERMET.

Deposition Parameters and resulting resistor values are summarized in Table
1.

\subsubsection{Homogeneity and yield}

The homogeneity of the CERMET resistors over a 256 strip detector is shown
in Fig. 9. Every 10th resistor was measured on a needle probe station. Non
of them exhibited a mal-function due to broken resistor lines or
insufficient contact performance. Obviously there is a left-right increase
in the resistor value with a small oscillation. So far it is not clear,
whether this comes from the magnetron sputtering profile or a varying
composition in the target. The average value was (2.98 $\pm $ 0.17) M$\Omega 
$.

\section{Conclusions}

A six mask process for fabricating AC-coupled GaAs microstrip detectors with
a new type of integrated biasing resistors was developed. Process
temperatures do not exceed 300${{}^{\circ }}$C in order to risk a
degeneration of the Schottky contacts.

Coupling capacitances with different dielectric layers have been integrated
onto microstrip detectors. For full size keystone detectors of 6 cm length
and 80 $\mu $m pitch, the corresponding values of the coupling capacitance
were 200 pF and 370 pF for SiO$_2$ and SiO$_2$/Si$_3$N$_4$, respectively.
The frequency dependence of the integrated capacitors is almost as good as
an external C-chip. Only a weak decrease of the capacitance above 100 kHz
was observed.

The measurements have shown that the electrical properties of the 2.47 M$%
\Omega $ CERMET resistors are quite reasonable and meet the basic
requirements. The sheet resistance of the 121 nm thick rf-sputtered layer at
100 W sputtering power is 78.4 k$\Omega $/sq.. It can be varied by the
sputtering power. The TCR of 2.5 $\times $ $10^{-3}/{{}^{\circ }}$C is
comparable to that of an external R-Chip that is frequently used for
nonintegrated microstrip detectors.

It has to be proofed, if the high strip resistance of 150 $\Omega $/cm
significant deteriorates the signal to noise level when amplifiers with fast
shaping times are used. The recently fabricated detectors are currently
being tested with respect to their charge collection efficiency and position
resolution.

\section{Acknowledgment}

This work has been supported by the BMFT under contract 057FR11I.

\newpage
\rule{0.00cm}{5cm}

\begin{center}
$
\begin{tabular}{llcc}
\multicolumn{4}{l}{Table 1: Deposition parameters and resistor values} \\ 
\hline
Sputtering power & [W] & 100 & 200 \\ \hline
Sputtering time & [s] & 250 & 100 \\ 
Film thickness & [nm] & 121 & 133 \\ 
Resistivity & [M$\Omega $] & 2.47 & 4.85 \\ 
Resistance & [$\Omega $cm] & 0.95 & 3.14 \\ 
Sheed resistance & [k$\Omega $/$\square $] & 78.4 & 236 \\ 
TCR & [$10^{-3}/{{}^{\circ }}$C] & 2.5 & 3.0 \\ \hline
\end{tabular}
$
\end{center}

\newpage
\epsfig{file=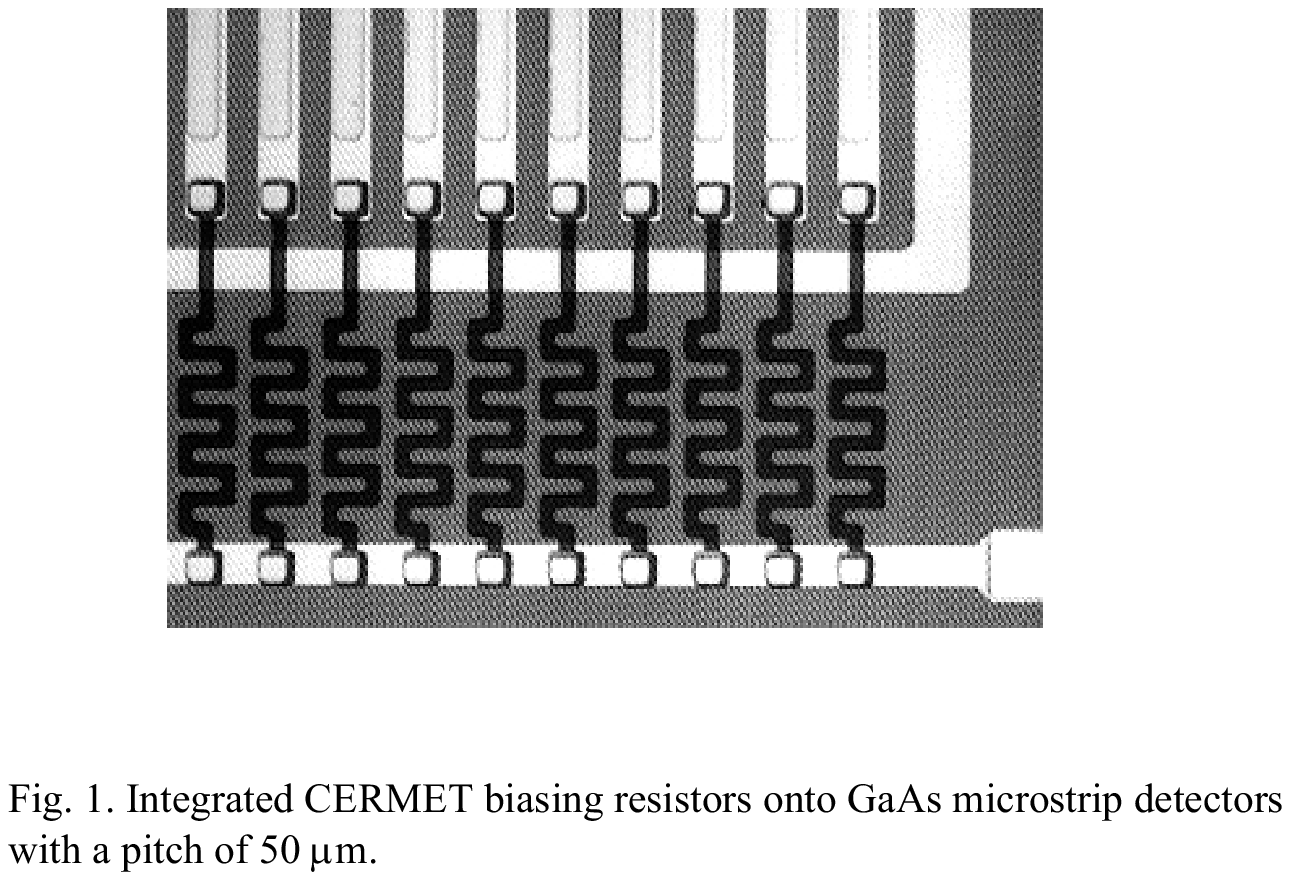,width=1.0\textwidth,height=20cm} 
\newpage
\epsfig{file=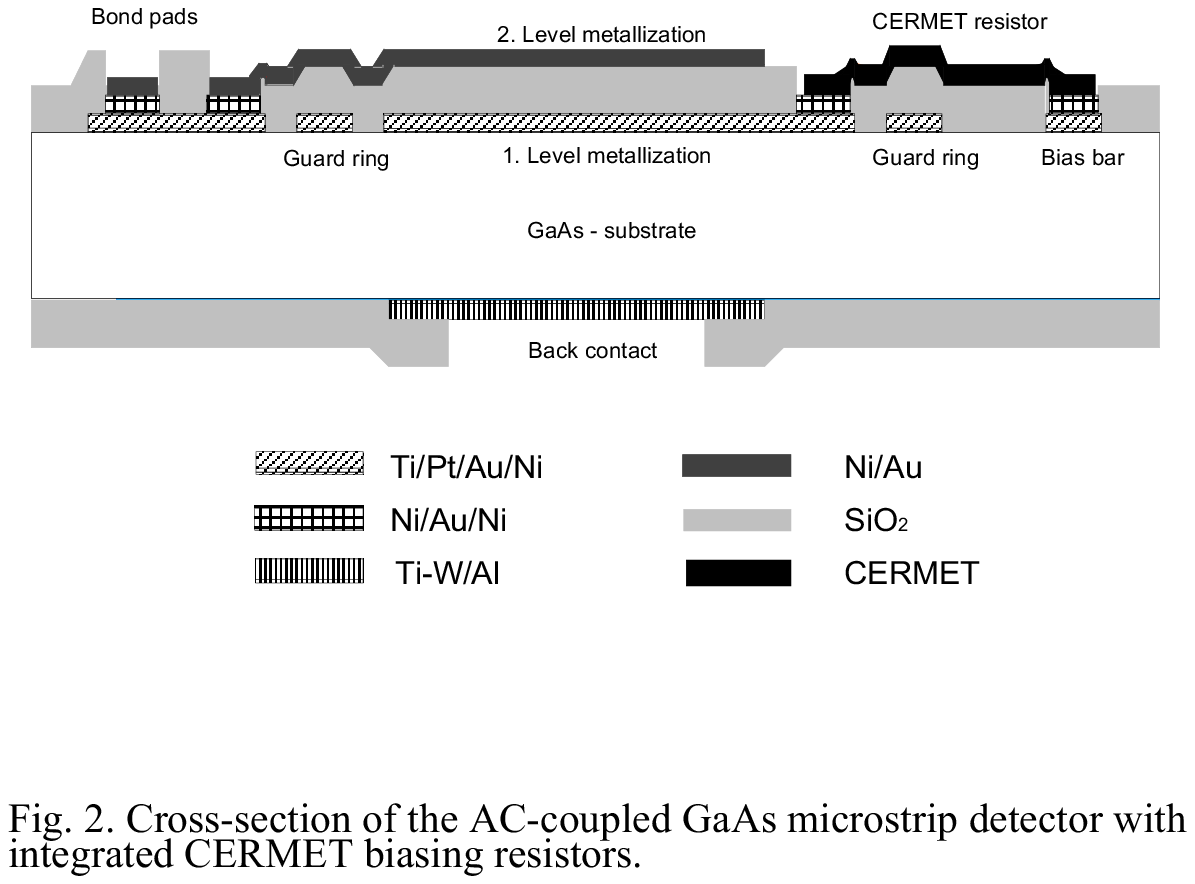,width=1.0\textwidth,height=20cm} 
\newpage
\epsfig{file=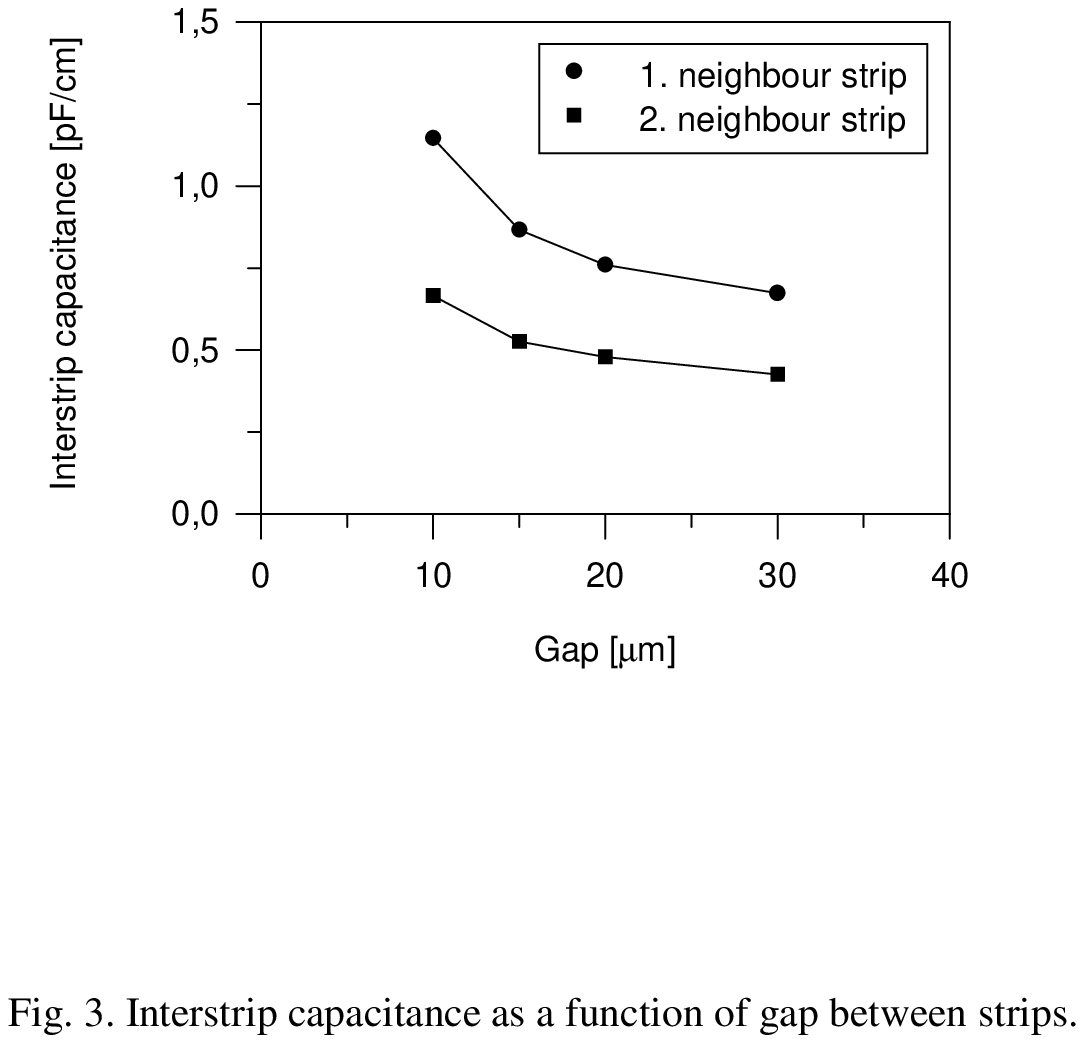,width=1.0\textwidth,height=20cm} 
\newpage
\epsfig{file=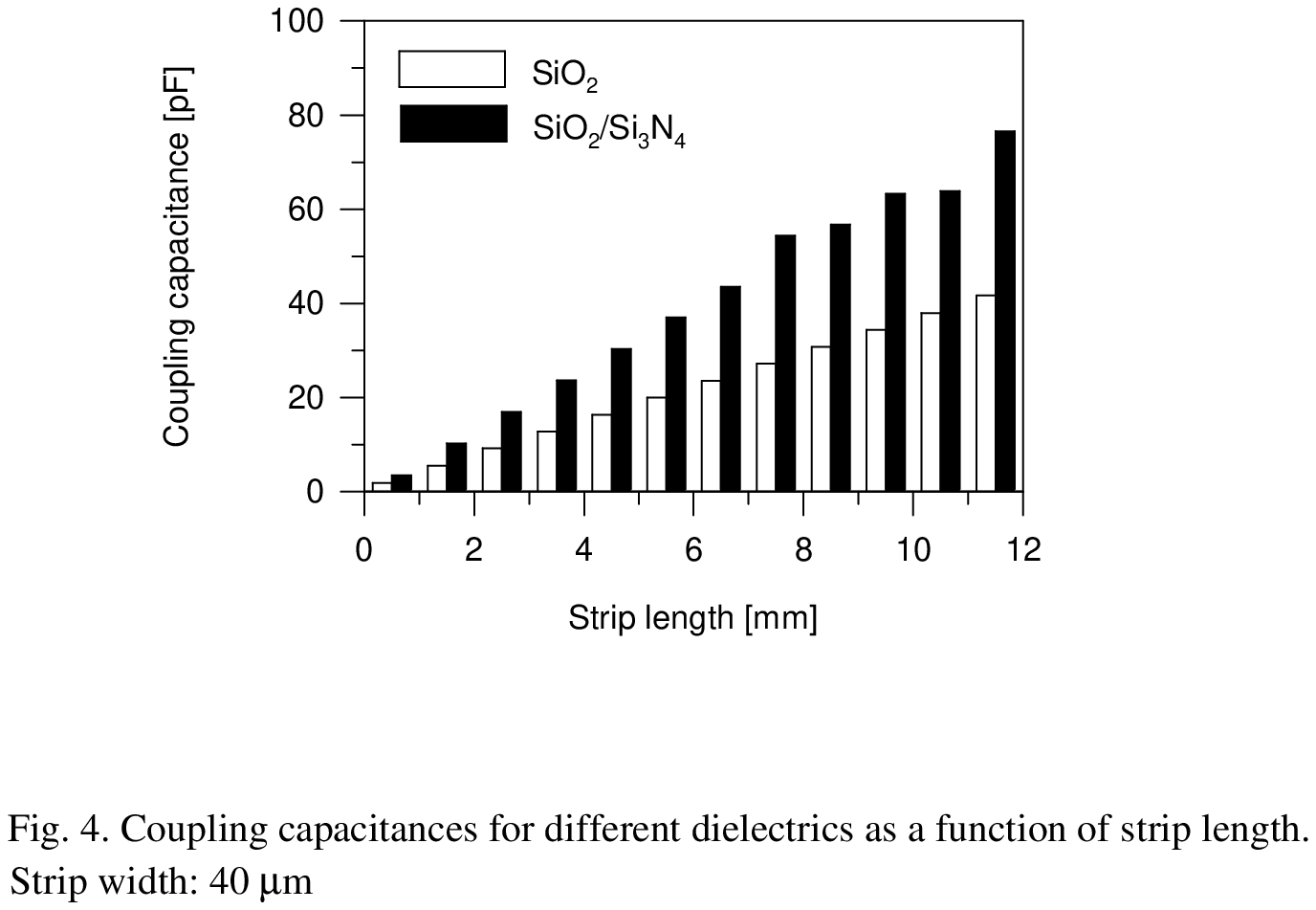,width=1.0\textwidth,height=20cm} 
\newpage
\epsfig{file=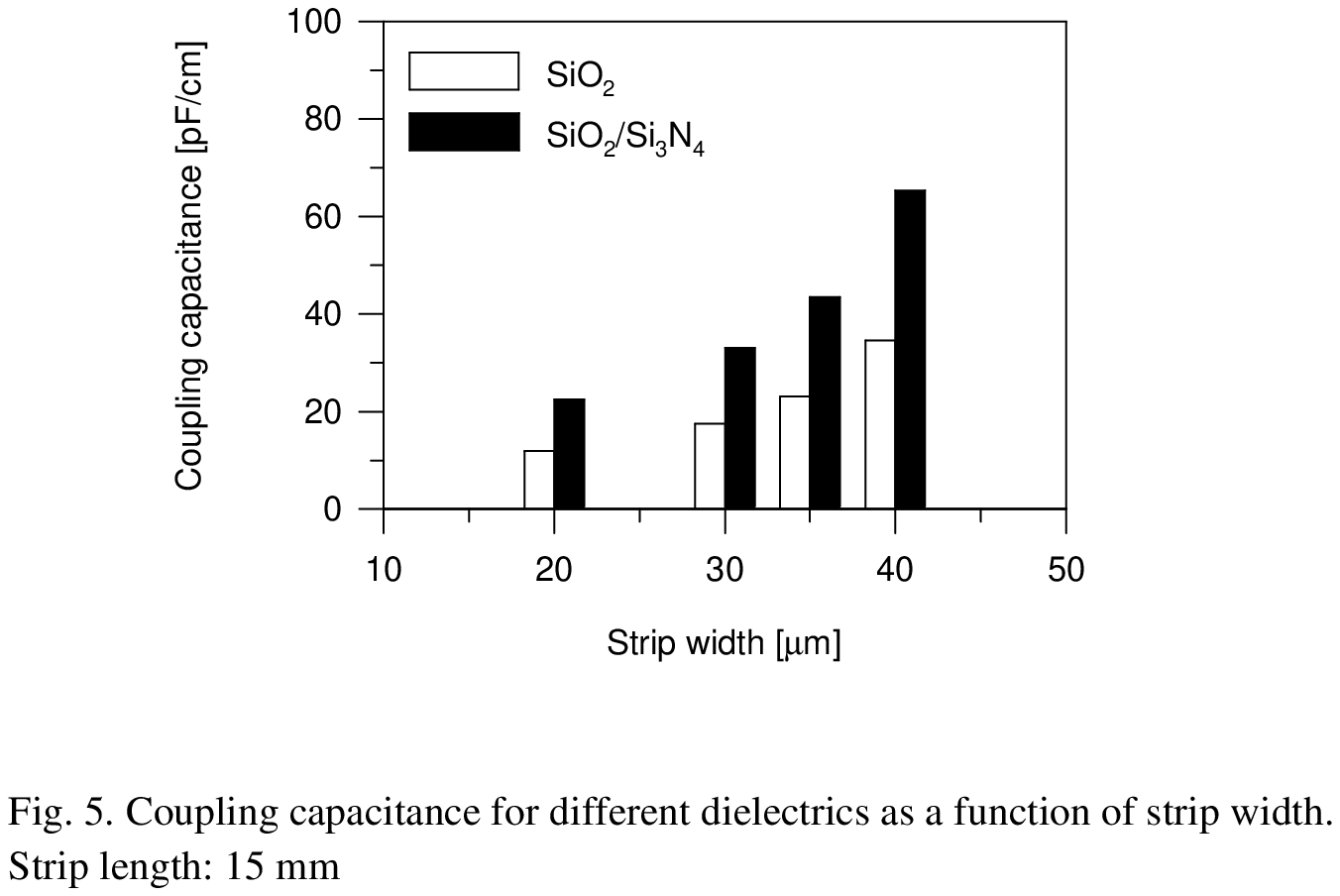,width=1.0\textwidth,height=20cm} 
\newpage
\epsfig{file=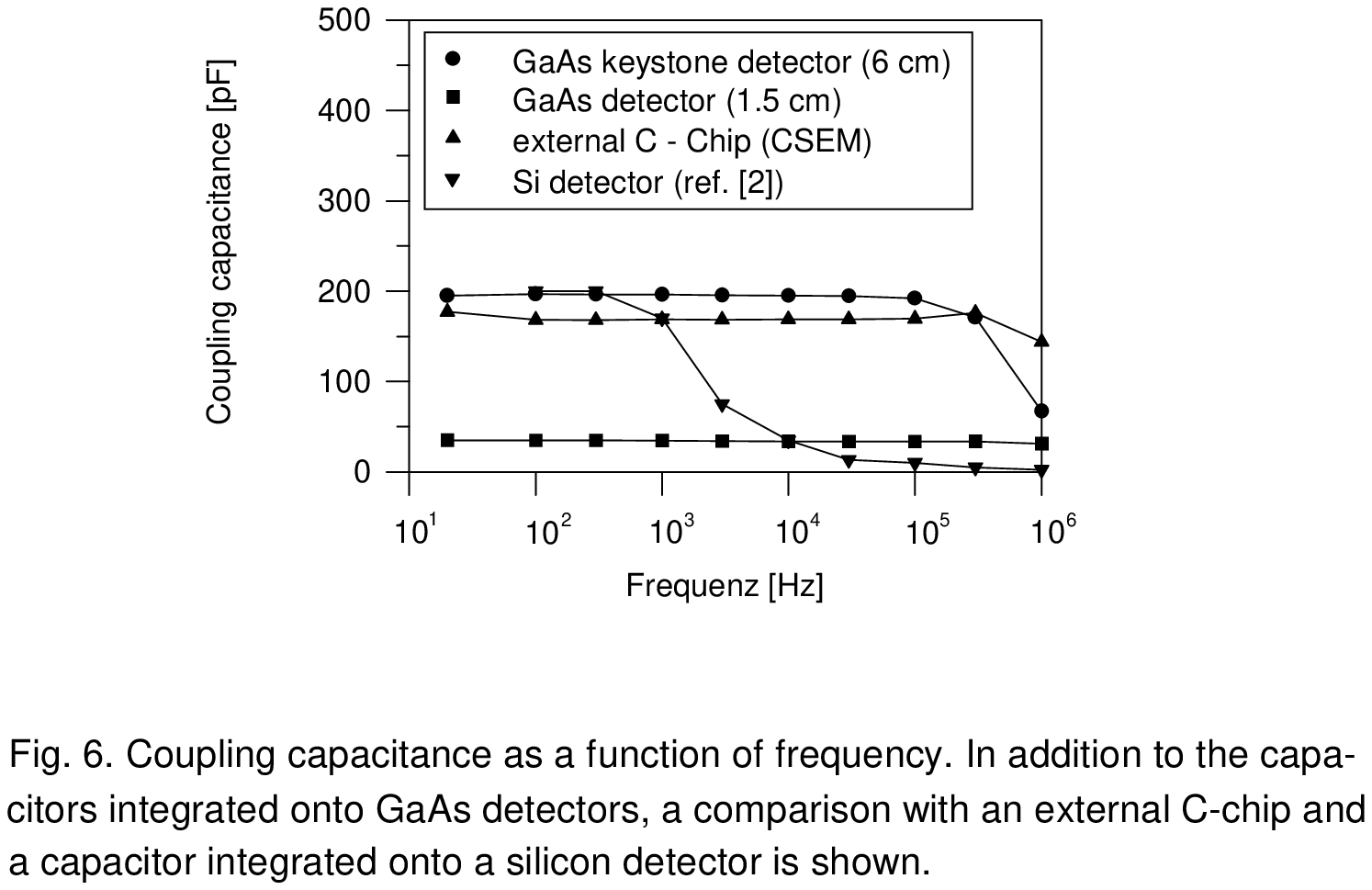,width=1.0\textwidth,height=20cm} 
\newpage
\epsfig{file=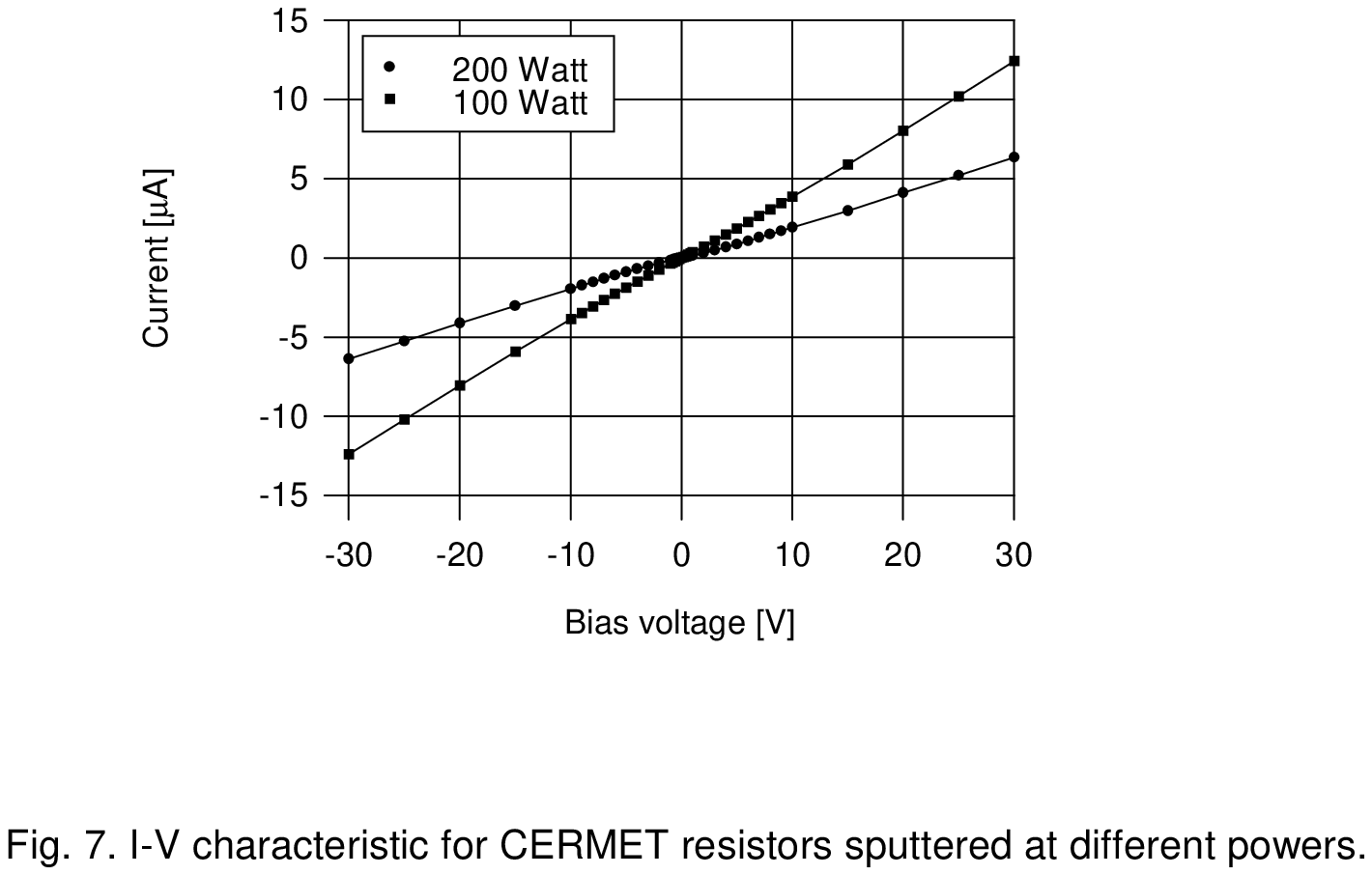,width=1.0\textwidth,height=20cm} 
\newpage
\epsfig{file=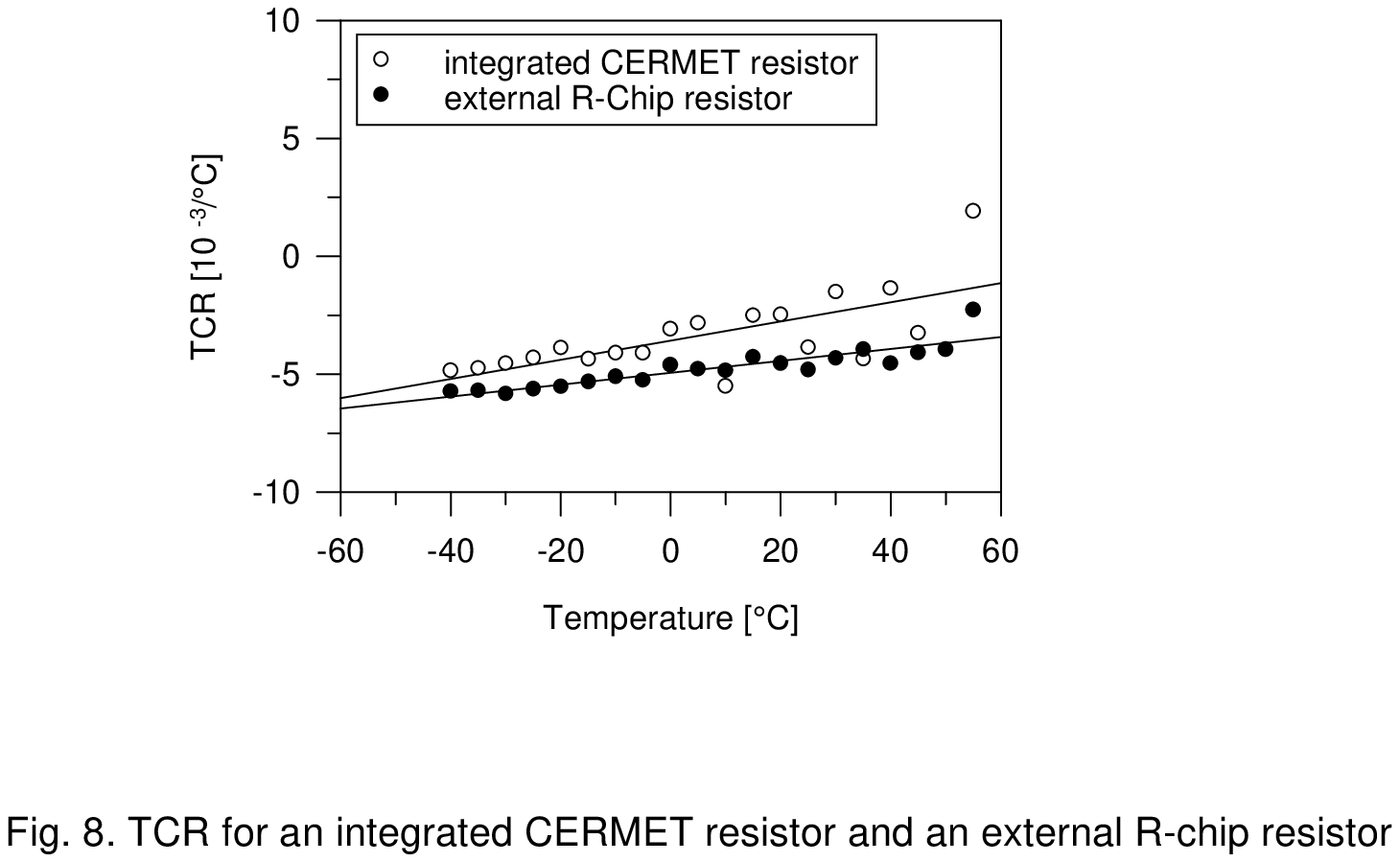,width=1.0\textwidth,height=20cm} 
\newpage
\epsfig{file=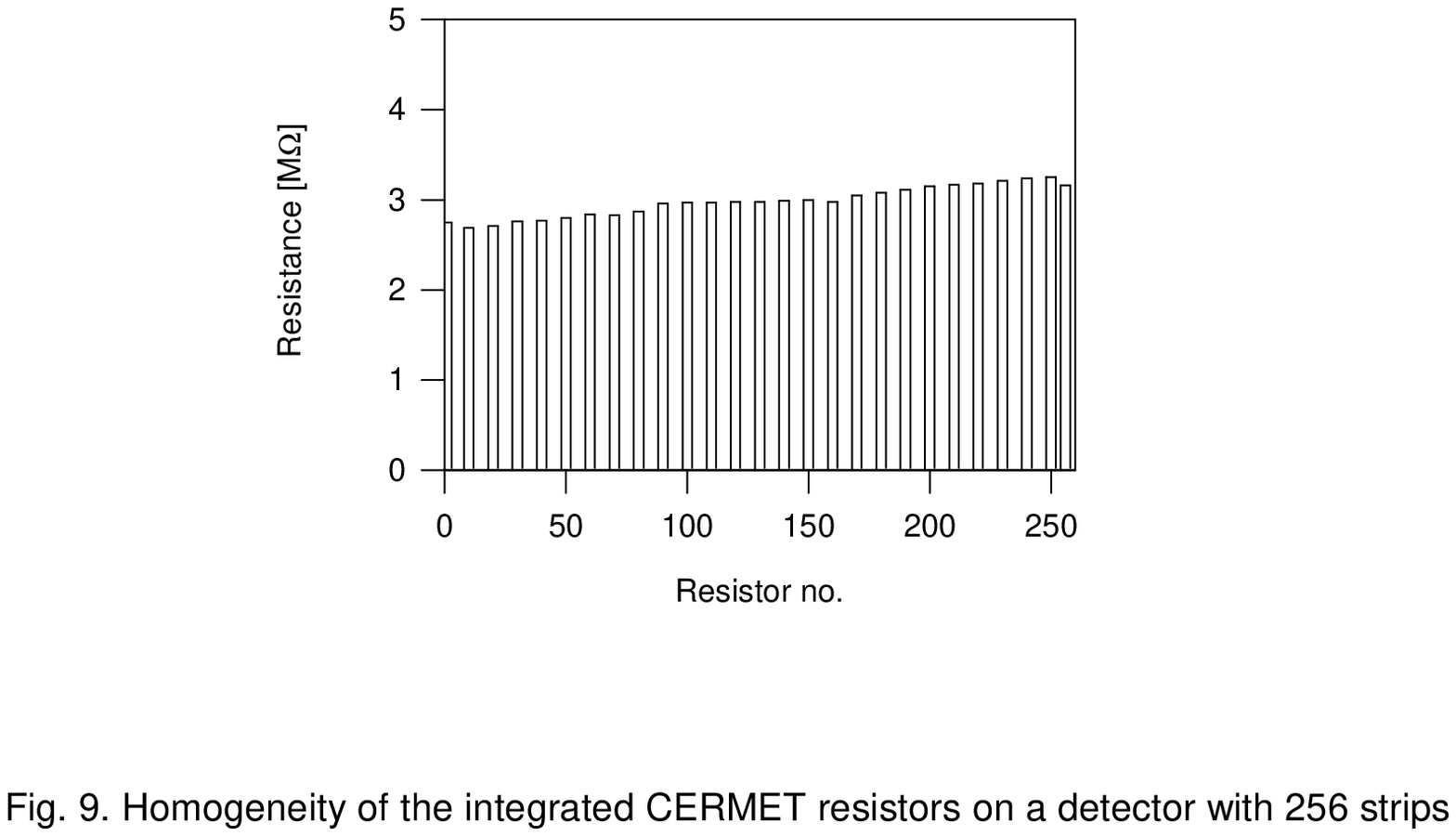,width=1.0\textwidth,height=20cm}

\end{document}